# Spatial and velocity offsets of Galactic masers from the centers of spiral arms

## Jacques P Vallée

National Research Council of Canada, Herzberg Astronomy & Astrophysics Research Center, 5071 West Saanich Road, Victoria, B.C., Canada V9E 2E7     jacques.p.vallee@gmail.com

**Abstract.**   Some galactic theories of spiral arms predict an offset between different tracers of star formation. Our goal here is to find such an offset between the observed location of radio masers and the location of the arms, using a recent 4-arm model fitted to the CO 1-0 gas.  Our method here is to compare a recent global 4-arm spiral model (as fitted to the arm tangents in the observed broad CO 1-0 gas) with the recent results for the trigonometric distances of radio masers, for the main arms (Cygnus-Norma, Perseus, Sagittarius-Carina, Scutum, and Norma).  Our results indicate that most radio masers are near the inner arm edge (toward the Galactic Center) of spiral arms. These masers are offsets from the model arm (where the broad CO 1-0 molecular region resides), by  0.34 ± 0.06 kpc inward. In radial velocity space, the median offset between masers and the CO-fitted model is around 10 ± 1 km/s. Based on the masers being observed here to be radially inward of the broad CO gas in the Cygnus arm at 15 kpc along the Galactic Meridian, the corotation radius of the Milky Way disk is >15 kpc distant from the Galactic Centre and the density-wave's angular pattern speed is <15 km/s/kpc. Arm pitch angle should be measured using many arm tracers, and located on both side of the Galactic Meridian, to ensure better precision and avoid a bias pertinent to a single tracer.

## 1. Introduction.

Looking from the Sun to each spiral arm towards the inner Galaxy, in Galactic Quadrants IV and I, radio and optical observations have already shown the precise galactic longitude of the peak intensity value, for each arm tracer (CO, dust, old stars. etc).  For each interior arm, some offsets between the dust lane and the peak intensity in CO 1-0 gas were observed, then culled into catalogues – see Table 3 in Vallée (2014a) and the Appendix in Vallée (2016a).

A typical picture shows a  compressed dust lane (located at the inner edge of a spiral arm) and the lower density CO 1-0 gas farther out (Fig.1 in Vallée 2014a; Fig.2 in Vallée 2014b).  This 'offset' pattern was also found to be mirror-imaged in Galactic longitudes (East and West of the Galactic Meridian) – see Fig. 2 in Vallée (2016a).

Can we observed these offsets anywhere else along the arms, and not just when the arms are seen tangentially from the Sun? The motivation for this study is to search for new physical offsets, closer to the Galactic Meridian.

To answer this question, we will make use of the recently published distances to radio masers, and compare them to the location of the CO 1-0 gas in spiral arms, as fitted to a recent 4-arm model. We will use statistics to infer the mean offsets for each arm.

The relevance of this study is important if we are to select a proper framework for spiral arms, and a proper theoretical model. The results for the Milky Way may be transposed later to other nearby spiral galaxies, given the proximity of the Milky Way's arms to our telescopes (better angular resolution, better tracer detectability, etc.). Also, these results may be employed later in the broader cosmological and galactic context.

Thus the density wave theory (DWT) of spiral arms predicts a physical offset between different tracers of star formation (location of young stars versus location of old stars); most other theories of spiral arms do not predict such a clear physical offset.

Some numerical arm models exist, such as the Besançon model (optical star counts, stellar density and stellar kinematics – see Bienaymé et al 2018) and the GalMod model (optical star counts – see Pasetto et al 2016; 2018). Their synthetic observations are numerical predictions, based on theoretical models, of the values that an astrophysical source should appear to an observer.

We have added some discussion on some recent numerical models, aiming at finding a comparison with our results (Besançon, GALMOD, Pasetto et al 2018, Sellwood et al 2019, Gaia Collaboration – Kats et al 2018, Roca-Fabrega et al 2013).

Here, we seek to find if the 'offset' pattern can be observed nearer the Galactic Meridian, using recently published trigonometric-measured locations for radio masers. Section 2 deals with the observed data and the chosen 4-arm model. Section 3 and 4 computed the offsets in physical and velocity spaces. Sections 5 and 6 present an interpretation of the offsets with respect to the arm tangents and to the density wave theory. A discussion (Section 7) follows on the ambiguous pitch angle values for various sections of an arm. The relevance of these new results to the DWT and GalMod models can be found in Section 8. This is followed by a conclusion (Section 9).

## 2. Data for the trigonometric masers, and recent spiral arm model

Radio trigonometry allows the precise measure of a distance, for nearby masers. **Tables 1 to 6** compile the results of a survey of the literature, for published data on radio masers in our Milky Way galaxy.

To avoid comparing observational data of different qualities, we restricted the data to those having a better precision. Thus in Tables 1 to 6, we excluded sources with a large parallax error >20% of the parallax value – here the fractional parallax error f = 0.20 or less.

Columns in these tables show the source name, galactic coordinates, distance (=1/parallax), distance range [1/(parallax + error); 1/(parallax – error)] distance error (half of the range), f value (fractional parallax error), systemic radial velocity and errors, and Reference to these data.

Error analysis. Estimating distances from parallaxes must be treated carefully – Bailer-Jones (2015) noted that objects with a fractional parallax error (f) larger than 20% require a bias correction. Since the parallax error is Gaussian, but not the error in distance, then a parallax-independent 'Prior' between 0 and 1 is employed to give the probability that the source is lying at the far distance; in the absence of such information a probability value of 0.5 is suggested (Reid et al 2016 – their Section 2.2; Zucker et al 2018 – their Section 3.1).

All of the maser data in Tables 1 to 6 here have a fractional parallax error f of 20% or less, hence they do not require a bias correction. A bias correction may be affected by the subjective 'Priors' chosen, and by the degree of validity of the assumptions chosen.

For the Perseus arm, the individual distance errors have a median symmetric distance error of only 0.20 kpc for 25 masers, while it is 0.29 kpc for 25 masers in the Sagittarius spiral arm.

To compare with a 4-arm model, we defined the mid-arm as follows. The mid-arm is populated by old stars and by the cold low-density CO 1-0 gas, so that a telescope scan in galactic longitudes shows a maximum in CO 1-0 intensity when crossing an arm, due to looking tangentially to that arm (through a longer distance along the long line of sight). The specific galactic longitude of the maximum CO 1-0 emissivity for each arm is then fitted to a 4-arm spiral model. That galactic longitude of maximum emissivity for CO 1-0 differs from the galactic longitude of that arm's dust lane, the CO 1-0 gas being farther away from the Galactic Center direction – see Fig. 2 in Vallée (2017b).

The dust arm location would correspond to the predicted location of the shock in the spiral density wave (as well as the gas density maximum). The mid-arm would then correspond roughly to the predicted location of the 'potential minimum' of the spiral density wave – see Fig. 2 in Roberts (1975).

So we employed a large-scale 4-arm spiral model, as fitted to the broad telescope scan in galactic longitudes, looking for the broad CO J=1-0 gas intensity obtained when each spiral arm is seen tangentially in that arm tracer. The search for the best fit spiral arm pitch angle, averaged over a large angular scale, employs the arm tangents on both sides of the Galactic Meridian (separating Galactic Quadrant I from Galactic Quadrant IV), using a single trigonometric equation (equation 10 in Vallée 2015). Thus the tangents from the Sun to a spiral arm, seen in Galactic Quadrant I at $l_I$ and seen in Galactic Quadrant IV at $l_{IV}$, will reveal the pitch angle p through ths equation:

$$\ln[\sin(l_I)/\sin(2\pi - l_{IV})] = (l_I - l_{IV} + \pi) \cdot \tan[p]$$

Here one should use the same arm tracer (CO, say) on both side of the Galactic Meridian. Other arm tracers (dust, say) will give very similar p values.

Only such a large-scale angular view can provide a precise global arm pitch angle value, and it can reveal the Milky Way's global order and its mirror-image symmetry across the Galactic Meridian (Vallée 2016a).

For the Scutum arm, seen tangentially at galactic longitudes 32° ±1 ° and 310° ±1°, one found a pitch angle of 13.3° ±1.0° – see Table 2 in Vallée (2015). For the Sagittarius arm, seen tangentially at galactic longitudes 50.5° ±0.5° and 283° ±1°, one found a pitch angle of 13.9° ±0.6°– see Table 1 in Vallée (2015). For the Norma arm, seen tangentially at galactic longitudes 18° ±2° and 329° ±1°, one found a pitch angle of 13.8° ±0.6°– see Table 1 in Vallée (2017b).

**Figure 1** shows the locations of the observed masers (filled squares) with respect to the arm middle (CO-fitted model curves) for the Cygnus (Outer Norma) arm in red (Table 1), the Perseus arm in yellow (data in Table 2 here), the Sagittarius arm in green (Table 3), and the Scutum arm in blue (Table 4), the Norma arm (Table 5), and the Outer Scutum arm in blue (Table 6). The small error bars for each maser are indicated in Tables 1 to 6 (columns 4 and 5).

The coloured curves were derived by fitting the galactic longitudes where the CO 1-0 gas had a maximum intensity as seen in a telescope with a beam of 8.8' scanning in galactic longitudes, yielding a specific galactic longitude for each spiral arm with an error in longitude of 0.1°. The parameters and uncertainties in the model are as follow. The spiral arm model employed a large-scale angular view to derive the pitch angle, as well as the latest known distance to the Galactic Center of 8 kpc ±0.1 kpc (Vallée 2017a). The start of each arm is also fitted to the available observations, yielding a radial distance to the Galactic Centre of 2.2 kpc ±0.05 kpc. It was fitted to the best galactic longitude of the arm tangent as observed in the broad CO J=1-0 gas (see Table 5 in Vallée 2016a). Also, it used the latest values for the orbital circular speed at 230 km/s ±3 km/s around the Galactic Centre (Vallée 2017a). Further description of the parameters of this global 4-arm spiral model were given in Vallée (2017b) and in Vallée (2017c).

Figure 1a shows the wider area (from the Outer Scutum blue arm at top, to the Outer Sagittarius green arm at bottom). Figure 1b shows the narrow area (from the Cygnus red arm to the Galactic Center).

### 3. Offsets of masers, from the arm middle (fitted to the CO 1-0 gas peak).

3.1 The Cygnus (Outer Norma) arm – red curve in Galactic Quadrants I to III
There are six trigonometric masers in the outer Galaxy (from Galactic longitudes 70° to 200°), and all have a small fractional parallax error < 14%. All masers appear in the inner edge of the arm, thus offset from the arm middle located at the broad CO 1-0 gas (shown by the arm model red curve).

3.2 The Perseus arm – yellow curve in Galactic Quadrants II and III
20 out of 25 masers are located inside of the mid-arm shown by the CO-peak curve; the offset of masers to CO gas is measured here at 0.31 ± 0.06 kpc (Table 7) and was measured earlier at 0.40 ± 0.10 kpc (Vallée 2018a).

3.3 The Sagittarius arm – green curve in Galactic Quadrants I and IV
20 out of 25 masers appear inside of the arm, offset from the arm middle at CO 1-0 gas (green curve). Statistics on the offset between masers and the CO green curve yields 0.38 ± 0.06 kpc (Table 7).
Wu et al (2019 – their Fig.7) showed 30 maser positions, of which 3 masers were beyond a galactic longitude of 49 degrees, but still within the tangent point for the CO 1-0 peak intensity, which was observed near a galactic longitude of 51.0° – see table 3 in Vallée (2016a).

3.4 The Scutum arm – blue curve in Galactic Quadrants I and IV

11 of the 15 masers are located on the interior of the arm (inner space), before the mid-arm (blue curve showing the CO 1-0 arm tangent). We find a mean offset of 0.19 ± 0.06 kpc (Table 7).

### 3.5 The Norma arm – red curve in Galactic Quadrant I

Only 3 trigonometric masers appear near the Norma spiral arm. 2 out of 3 masers appear inside of the arm, offset from the arm middle at the broad CO 1-0 arm tracer (red curve). We find a mean offset near 0.12 ± 0.06 kpc (Table 7).

### 3.6 The Outer Scutum arm beyond the Galactic Center – blue curve in Galactic Quadrants I and II

Sanna et al (2017) found the G007.47+00.05 maser at a parallax distance of 20.4 kpc, with a large error bar. Within this large error bar, one could assign it to a possible extension of the Scutum arm beyond the Galactic Center.

However, as the arm width should be less than 700 pc, then one could not separate this imprecise maser location with any arm features (inner, center, outer). That maser (Table 6) could well be inside the arm at the broad CO 1-0 arm (blue curve), within its large observational error.

## 4. The masers as displayed in velocity space.

**Figure 2** shows the masers (filled squares) and the 4-arm model (curves), on a plot of radial velocity versus galactic longitude, covering the area towards the Galactic Center (Fig. 2a) and away from it (Fig. 2b). It is clear that the masers as spatially identified near the spiral arms (in Fig. 1) are also kinematically identified near their arms (Fig. 2). The small error bars for each maser are indicated in Tables 1 to 6 (column 7).

In the Sagittarius arm and in the Scutum arm within Galactic Quadrant I (Fig. 2a), the masers are located at a slightly higher positive radial velocity than the CO model (by about 10 to 30 km/s), clearly indicative of their spatial locations closer to the Galactic Center in Fig. 1. Masers closer to the Galactic Center than the CO-fitted arm position would show a more positive radial velocity, because the radial velocity gradient at galactic longitude $20°$ is known to be +145 km/s for the first 7.5 kpc from the Sun (Fig. 2a in Vallée 2008 for galactic longitude $20°$), while the orbital velocity gradient predicted by a density-wave shock is 20 km/s over 0.5 kpc in the arm (Fig. 12 in Roberts (1975).

In the Perseus arm and Cygnus arm within Galactic Quadrant II (Fig. 2b), the masers are located at a slightly higher negative radial velocity than the CO model (by about 10 to 20 km/s). Masers farther away from the Galactic Center than the CO-fitted arm position would show a more negative radial velocity, because the radial velocity gradient at galactic longitude 160° is known to be -32 km/s over the first 7.5 kpc from the Sun (Fig. 2b in Vallée 2008 for galactic longitude 160°), while the orbital velocity gradient predicted by a density-wave shock is 20 km/s over 0.5 kpc in the arm (Fig. 12 in Roberts (1975).

We note that the high-velocity maser in Norma (+108 km/s at longitude 12 degrees) belongs to the beginning of the Inner Norma arm. The maser near the Outer Scutum arm (-16 km/s at galactic longitude 7.5 degrees) may be closer in velocity space to the Outer Sagittarius arm (within a large distance error bar).

## 5. The various arm tangents and their offsets

The various trigonometric masers are not easily aligned along a narrow arc, but cover a rather large range in galactic longitude (seen in Figure 1). The broad CO J=1-0 emission peak is an arm tracer that is rather narrower in galactic longitude. The spiral arm width is about 700 pc, as found in Galactic Quadrants IV and I (Fig. 2 in Vallée 2016a).

For the Sagittarius arm, observations showed that the broad CO is at a median longitude = 51.0° (Table 5 in Vallée 2016a). Within 10° (about 730 pc) of that value, the observed masers have a median longitude of 49.2° (from Table 3 here), giving an inward offset of 1.8° ±0.1° (equal to 130 pc ±7 pc, at a distance of 4.2 kpc). Observations showed that the 870 µm dust is at a median longitude of 49.1° (Table 3 in Vallée 2016a).

**Figure 3** shows these masers as a function of galactic longitudes, for the Sagittarius arm (Fig. 3a) and for the Scutum arm (Fig. 3b).

For the Scutum arm, observations showed that the broad CO is at a median longitude = 33.5° (Table 5 in Vallée 2016a). Within 7° (about 710 pc) of that value, the masers have a median longitude of 30.6° (from Table 4 here), giving an inward offset of 2.9° ±0.1° (equal to 290 pc 10 pc, at a distance of 5.8 kpc). Observations showed that the 870 µm dust is at a median longitude of 30.9° (Table 3 in Vallée 2016a).

**Figure 4** shows these masers as a function of the offset (distance) from the nearest 4-arm model (itself fitted to the CO 1-0 arm tangents in galactic longitude). This offset is measured linearly on Figure 1, starting from the maser location and going to the nearest arm. These masers are not aligned along a narrow arc, but possess a large trail in galactic longitude (several tens of degrees – see Fig. 3) and a large offset from the CO 1-0 arm tracer (close to 1 kpc – see Fig. 4).

### 6. Offset statistics and the density wave predictions.

Observations with a telescope made along galactic longitudes eventually reach the tangent to a spiral arm, recording the peak intensity in a tracer (dust, say), thus finding the exact galactic longitude where that tracer has shown a peak in intensity in a spiral arm. Observations made with another tracer (CO 1-0, say) along galactic longitudes will record the maximum intensity at a slightly different galactic longitude, for the same spiral arm. There is an offset in galactic longitudes between the two tracers, with the dust peaking at a lower galactic longitude than the CO 1-0 gas in Galactic Quadrant I, and the CO 1-0 gas peaking at a lower galactic longitude than the dust in Galactic Quadrant IV (a mirror image is created – see Fig. 4 in Vallée, 2017d).

Here, the offset between a maser versus the CO-fitted arm is measured to the spiral arm curve in Figure 1.  **Table 7** shows in column 3 the spatial offsets between the masers and the mid-arm.  The median offset value found in Table 7 is 0.34 ±0.06 kpc.

The density wave model is the major arm model predicting such offsets between the compressed dust lane, maser region, young stars, low-density broad CO gas intensity integrated over a long distance along the line of sight (at the arm tangent), and old stars (Roberts, 1975), as was observed near the arm tangents (see the offset between the red zone and the blue zone in Fig. 4 in Vallée 2017d).

The density wave model of Roberts (1975 – his Fig.2) predicts an offset between the shock/dust (maser lane), and the 'potential minimum' (CO gas peak) of 465 pc (3.7% of an arm cycle, with 4 arms, at a galactic radius of 8.0 kpc). The model of Gittins & Clarke (2004 – their Fig.11) predicts an offset of 402 pc (3.2% of an arm cycle), while the model of Dobbs & Pringle (2010 – their Fig. 4a) predicts an offset of 352 pc (2.8% of an arm cycle).

Co-rotation - This offset pattern is predicted by the density-wave, when the arm is closer to the Galactic Center than the 'co-rotation radius'.  It follows that both the Perseus arm (near 11 kpc on the Galactic Meridian) and the Cygnus arm (near 15 kpc on the Galactic Meridian) are closer to the Galactic Center than the co-rotation radius ($R_{coro}$ >15 kpc). For a circular orbital gas velocity of 230 km/s, this implies an angular speed of the density-wave pattern speed $\Omega_p$ < 15 km/s/kpc – this makes the Milky Way's $\Omega_p$ similar to that for the galaxy NGC 2997 ($\Omega_p \approx 16$ km/s/kpc) as found by Ghosh and Jog (2016) and similar to the prediction for the outer Milky Way galaxy ($\Omega_p \approx 16$ km/s/kpc) by Foster and Cooper (2010 – their Section 3.2.2).

The predictions of the density-wave theory for the co-rotation radius are often made for a spiral galaxy in isolation. These predictions could shift if there is a nearby dwarf galaxy near the Cygnus arm or Perseus arm (Olano 2016; Laporte et al 2018) or near the 'Local arm' (Vallée, 2018b).

**Table 8** shows the radial velocity offsets between the masers and the CO-fitted arm model. The median offset value in radial velocity is 10 ± 1 km/s. The density wave model predicts an offset between the shock/dust (maser lane), and the 'potential minimum' (CO gas peak) of 20 km/s – see Fig. 12 in Roberts (1975) for their idealized galaxy.

## 7. Ambiguous arm pitch angle values, from radio masers alone

Global arm model. Our global spiral arm model employs an arm's pitch angle obtained over a global view in galactic longitude, encompassing two galactic quadrants across the Galactic Meridian, one to the left and one to the right side. In addition, our global view of the arms' pitch angle employed different arm tracers (dust, CO 1-0, etc), finding very similar values (within 1° or so) – see Section 2.

Segmented arm models. Below here, we analyse other results for an arm's pitch angle value, as obtained over a smaller view in galactic longitude, using a single side of the Galactic Meridian, and often with 'gaps' in galactic longitude (models taking only a few arm segments). In addition, most radio papers reporting on segmented arm pitch values have employed only masers as arm tracers (excluding information from other arm tracers such as dust, CO 1-0, etc). A segmented view may differ artificially from a global view, owing to an inherent incompleteness.

The published pitch angle values, using a limited arm segment and employing trigonometric masers, vary a lot, notably for the Sagittarius arm (-7.2° in Wu et al 2019; -13.5° in Xu et al 2018; -19.0° in Krishnan et al 2017; -6.9° in Reid et al 2014; -11.2° in Sato et al 2010), and for the Scutum arm (-18.7° in Xu et al 2018; -19.2° in Krishnan et al 2015; -7.0° in Reid 2012), or else for the Perseus arm (-9.0° in Zhang et al 2019; -11.1° in Sakai et al 2015; -9.4° in Reid et al

2014; -9.9° in Choi et al 2014; -9.5° in Zhang et al 2013; -17.8° in Sakai et al 2012; -13.0° in Reid 2012; -16.5° in Reid et al 2009).

Thus one reads a difference of about 10° between these maser-derived values within one arm, yet their pitch angle errors are typically listed as only 1°. How can that be? Some masers may not be in the arm proper (interarm, spur), or have too large a distance error (improper arm fit).

Interarm masers - A look at Figure 1 and Figure 3 suggests that some trigonometric masers may be closer to the middle of an interarm than to a spiral arm. If so, those interarm masers should *not* be employed to derive the pitch angle of the arm proper; only masers in the arm should be used to find the arm's pitch angle. This alleged variation in pitch angle over different segments of the same arm suggests that one cannot extrapolate the pitch of an arm segment in order to get the pitch of the whole arm (see table 1 in Vallée 2017c). Also, it appears that some arm segments lack masers, like the Perseus arm between galactic longitude 50° to 80° with very few masers (Zhang et al 2013).

Masers in spurs – Some star formation (including masers) can happen in a small spur within an interarm, notably around the Sun (Vallée 2018b) and elsewhere. Thus one such interarm spur or loop has been observed, covering galactic longitudes 32° to 39° and radial velocities covering from +60 km/s to +90 km/s, located in between the Sagittarius arm and the Scutum arm (Rigby et al 2016). These spurs or loops may not necessarily be linked to large spiral arms.

Distant masers – The measured trigonometric distance to a distant maser comes with a larger distance error. With an increasing distance error at larger distances, it becomes problematic to fit an arm pitch, giving a larger error in arm pitch values. Thus models using a local segment of an arm, not a whole arm, to extrapolate a global arm pitch (with a large error), often result in numerous alternative positions of the same arm as extrapolated at a large distances (e.g., Fig. 2 in Sanna et al 2017). A small distance error near the Sun and a large distance error near the Galactic Center can allow a whole range in pitch angle.

Thus, trying to get a pitch angle value for an arm, using trigonometric masers alone (neglecting other arm tracers like CO 1-0, dust, etc; neglecting tracers from the same arm in the opposite galactic quadrant) incorporates the following problems. In finding the arm's overall pitch angle, one should question the presence of some masers (in interarm locations, in interarm spurs, in misaligned arm segments) and those with a larger distance error bar. These complexities should warrant larger published error bars in pitch angle values (not just 1°). Arm pitch angle values should be measured using many arm tracers, and located on both side of the Galactic Meridian, to ensure a better precision and to avoid a bias pertinent to a single tracer. The segmented spiral arm model of Reid (2017) and Reid et al (2016 – their Fig.1) roughly covers one-half of the Milky Way (Galactic Quadrants I and II), based on masers observed trigonometrically and primarily in these quadrants. Also, these segmented arm models often excluded observational data obtained with other tracers (gas, dust, etc).

A similar argumentation could be made for the so-called 'arm width' when computed from the width of the radio masers alone in the arm; the rather small arm width value (0.30 kpc in Sakai et al 2019) does not encompass the other arm tracers (dust, CO 1-0, old stars, etc). The arm width was found elsewhere to be near 0.7 kpc - twice the width of the dust offset to the mid arm of old stars and low-density CO 1-0 gas (as averaged over a long tangent line of sight from the Sun) – see Fig. 2 in Vallée 2016a). The real arm width must be inclusive of all arm tracers (Vallée 2016a), thus not just the width where the masers are located in the inner side of an arm (Reid et al 2016).

## 8. Relevance of these new results to existing numerical models

The self-consistent numerical Besançon galaxy model [BGM] of Bienaymé et al (2018) includes an axisymmetric gravitational potential; this model does not include spiral arms. Our own results here of an offset among start forming tracers cannot be compared to the Besançon model without arms.

The GalMod galaxy model of Pasetto et al (2018) includes an asymmetric bar and spiral arms; in addition, several profiles can be employed for the star formation rate (constant, exponential, linear, complex), and the locations of the spiral arms are logarithmic with a pitch angle of $8°$, with 2 arms. The earlier model of Pasetto et al (2016) also used 2 arms with a $8°$ pitch angle. Our own results here are based on a pitch angle near $13°$ for each of the 4 spiral arms. Their current GalMod model does not starts star formation when orbiting gas reach a shocked/dust lane, like in the density-wave model, so they do not predict an offset among star formation tracers.

The numerical models above choose only one bar near the center of the Galaxy, often with a large radius. Observations presented elsewhere noted 3 bars, each of different length and mass (Vallée, 2016b – his tables 2 and 3 and figure 3); the short boxy bar of radius 2.1 kpc has been shown to be the most physically massive.

On the dynamical side, the model of Sellwod et al (2019) explores the radial velocities of nearby stars from the Gaia DR2 catalogue, in order to test several theories of spiral structure; they favour the transient spiral mode with a strong inner Lindblad resonance, employing m=3 arms with a pitch angle near 26°, and an Inner Lindblad resonance of 8.1 kpc near the Sun. On the observational side, the corotation radius has been found to exceed 11 kpc for the Milky Way (Vallée 2018a).

The extensive structure observed by Gaia in the local kinematics of stars may have an external cause. The Local Arm feature near the Sun has been studied in many tracers already, with different physical pictures having emerged in its extent, shape, pitch and origin (for a review, see Vallée 2018b). The least controversial model involved the import of mass from elsewhere (grabbing a portion of the nearby Sagittarius dwarf galaxy at each of its successive passage near the Sun); the imported mass is later deformed by the differential rotation of the Milky Way into becoming elongated parallel to the existing spiral arms (Vallée 2018b). Thus the kinematics of all the nearby stars will be affected, including by the passage of galactic tides and of density waves, by nearby arms, by central bars, and by dark matter sub-halos (Gaia collaboration 2018).

The Gaia DR2 employed one billion stars and deduced a mean azimuthal velocity of 230 ±2 km/s near the Sun (Fig 13a in Gaia Collaboration 2018) – excluding stars higher than 0.6 kpc in galactic latitudes. This value is 5-sigma away from the value of 240 proposed by the Reid (2017) model.

The Gaia DR2 data used the 2-arm Drimmel (2000) model (Fig. 19 in Gaia Collaboration 2018); one arm of that model is close to the Scutum arm (but with an arm tangent seen from the Sun at a wrong galactic longitude), while the other arm is halfway between the Perseus arm and the Cygnus arm (beyond the Perseus arm) in the 4-arm model.

To explain some dynamical signature of the stars near the Sun, the bar model of Monari et al 2017a; 2017b) included a fast short bar, excluding a long bar (of radius of 5 kpc) to predict the bimodality of the local velocity space.

Some numerical models only use a single type of particle to represent stars, but nothing else to represent gas (CO) or dust. As such, our results with offsets, between (old stars and low-density CO gas) versus (young stars and dust lane), cannot be compared to these simple numerical models (e.g., Roca-Fabrega et al 2013, Antoja et al 2016, etc).

## 9. Conclusion

Here we collated all masers with a known trigonometric distance, excluding those with a large fractional parallax error (Tables 1 to 6). Then we employed the 4-arm model of Vallée (2017b; 2017c), as fitted to the arm tangents seen in CO 1-0 gas in Galactic Quadrants I and IV. Our global 4-arm model employs the arm tangents on both sides of the Galactic Meridian, for each arm. We computed the offsets of masers from that of the low-density CO 1-0 gas in a long line of sight through the arm (0.34 ±0.06 kpc and 10 ±1 km/s, in Tables 7 and 8).

Here we confirm that there is an 'offset' pattern for each arm. The trigonometric masers are mostly found closer to the Galactic Center than the CO 1-0 arm tracer.

Our results are as follow:
1. The Cygnus arm, beyond the Perseus arm, has all of its 6 radio masers located on its inner arm side (towards the Galactic Center) – see Fig.1.
2. As found earlier (Vallée 2018a), it is confirmed here that the Perseus arm has most of its radio masers located on its inner arm side – see Fig.1.
3. For the first time, it is found that the Sagittarius arm has most of its radio masers located on its inner arm side – see Fig.1.
4. The Scutum arm has most of its radio masers located on its inner arm side – see Fig. 1.
5. Also, the Norma arm has most of its radio masers located on its inner arm side – see Fig. 1.
6. We found offsets in velocity space between the masers and the 4-arm model fitted to the broad CO tangents, compatible with the density-wave model with shocks – see Fig.2.
7. We found a large width for the trail of masers in the Sagittarius and the Scutum spiral arms, as seen in galactic longitudes (Fig. 3).
8. We found the physical offsets of the masers, away from the CO 1-0 tracers taken at the mid-arm, with a median offset at 0.34 kpc towards the Galactic Center (Fig. 4).

Some comparisons with the predictions of density-wave theory would push the value of the co-rotation radius outward, nearer 15 kpc from the Galactic Center (Section 6). Finally, the error in the pitch angle value, made using various segmented arm models, is discussed above (Section 7); segmented arm models do not compare well to a global model where a smaller error in pitch angle is obtained using a wide-angle view (two galactic quadrants at once).

**Acknowledgements**

The figure production made use of the PGPLOT software at NRC Canada in Victoria. I thank an anonymous referee for useful, careful, insightful and historical suggestions.

**Table 1. Trigonometric masers near the Cygnus (Outer Norma) spiral arm (red curve).**

| Name | Gal. Long. (o) | Gal. Lat. (o) | Distance D [range] (kpc) | Half range (kpc) | Fract. parall. error (f) | Syst. V$_{lsr}$ (km/s) | Reference |
|---|---|---|---|---|---|---|---|
| G075.29+01.32 | 075.3 | +1.3 | 9.26 [8.85-9.71] | ±0.43 | 0.05 | -58 ±5 | Reid et al (2014) |
| G097.53+03.18 | 097.5 | +3.2 | 7.52 [6.66-8.62] | ±0.08 | 0.13 | -73 ±5 | Reid et al (2014) |
| G135.27+02.79 | 135.3 | +2.8 | 5.99 [5.62-6.41] | ±0.40 | 0.07 | -72 ±3 | Reid et al (2014) |
| G168.06+00.82 | 168.1 | +0.8 | 4.98 [4.44-5.65] | ±0.60 | 0.12 | -28 ±5 | Hachisuka et al (2015) |
| G182.67-03.26 | 182.7 | -3.3 | 6.71 [6.25-7.24] | ±0.50 | 0.07 | -07 ±10 | Reid et al (2014) |
| G196.45-01.67 | 196.4 | -1.7 | 5.29 [4.98-5.65] | ±0.34 | 0.06 | +19 ±5 | Reid et al (2014) |

Note : Taking all trigonometric maser sources, identified as located in the 'Cygnus' (outer Norma) arm in Reid et al (2014) or Hachisuka et al (2015), and some others, excluding sources with a large fractional parallax error f > 0.2. All distances are trigonometric; the published parallax (p, in mas) was converted to a distance (D, in kpc) through the equation $D = 1/p$.

==============================================================

**Table 2. Trigonometric masers near the Perseus spiral arm (yellow curve)**

| Name | Gal. Long. (o) | Gal. Lat. (o) | Distance D [range] (kpc) | Half range (kpc) | Fract. parall. error | Syst. V$_{lsr}$ (km/s) | Reference |
|---|---|---|---|---|---|---|---|
| G043.16+0.01 | 043.2 | +0.0 | 11.11 [10.42-11.90] | ±0.74 | 0.07 | +11 ±5 | Zhang et al (2019) |
| G048.60+0.02 | 048.6 | +0.0 | 10.75 [10.20-11.36] | ±0.58 | 0.05 | +18 ±5 | Zhang et al (2019) |
| G070.18+1.74 | 070.2 | +1.7 | 7.35 [6.67-8.20] | ±0.76 | 0.10 | -23 ±5 | Zhang et al (2019) |
| G100.37-3.57 | 100.4 | -3.6 | 3.46 [3.28-3.66] | ±0.19 | 0.06 | -37 ±10 | Choi et al (2014) |
| G108.20+0.58 | 108.2 | +0.6 | 4.41 [3.79-5.26] | ±0.74 | 0.16 | -49 ±5 | Choi et al (2014) |
| G108.47-2.81 | 108.5 | -2.8 | 3.24 [3.13-3.34] | ±0.10 | 0.03 | -54 ±5 | Choi et al (2014) |
| G108.59+0.49 | 108.6 | +0.5 | 2.47 [2.28-2.69] | ±0.20 | 0.08 | -52 ±5 | Choi et al (2014) |
| G111.25-0.76 | 111.3 | -0.8 | 3.79 [3.52-4.10] | ±0.29 | 0.08 | -40 ±5 | Sakai et al (2019) |
| G111.25-0.77 | 111.3 | -0.8 | 3.34 [3.11-3.61] | ±0.25 | 0.07 | -43 ±5 | Choi et al (2014) |
| NGC 7538 | 111.5 | -0.8 | 2.65 [2.54-2.77] | ±0.12 | 0.04 | -57 ±5 | Choi et al (2014) |
| IRAS 00420 | 122.0 | -7.1 | 2.13 [2.04-2.22] | ±0.09 | 0.04 | -44 ±5 | Reid et al (2009) |
| NGC 281 | 123.1 | -6.3 | 2.82 [2.60-3.08] | ±0.24 | 0.08 | -31 ±5 | Reid et al (2009) |
| W3(OH) | 134.0 | +1.1 | 1.95 [1.92-1.99] | ±0.04 | 0.02 | -45 ±3 | Reid et al (2009) |
| S Per | 134.6 | -2.2 | 2.42 [2.33-2.53] | ±0.11 | 0.04 | -39 ±5 | Choi et al (2014) |
| IRAS05168+36 | 170.7 | -0.2 | 1.88 [1.71-2.09] | ±0.19 | 0.10 | -19 ±3 | Sakai et al (2012) |
| G173.48+2.44 | 173.5 | +2.4 | 1.68 [1.64-1.72] | ±0.04 | 0.02 | -12 ±5 | Sakai et al (2019) |

| Name | Gal. Long. (o) | Gal. Lat. (o) | Distance D [range] (kpc) | Half range (kpc) | Fract. parall. error | Syst. $V_{lsr}$ (km/s) | Reference |
|---|---|---|---|---|---|---|---|
| G183.72-3.66 | 183.7 | -3.7 | 1.59 [1.56-1.62] | ±0.03 | 0.02 | +3 ±5 | Choi et al (2014) |
| G188.79+1.03 | 188.8 | +1.0 | 2.02 [1.67-2.54] | ±0.44 | 0.20 | -5 ±5 | Reid et al (2014) |
| Sharpless 252 | 188.9 | +0.9 | 2.10 [2.07-2.17] | ±0.03 | 0.02 | +8 ±4 | Choi et al (2014) |
| G188.94+0.88 | 188.9 | +0.9 | 2.15 [1.97-2.36] | ±0.20 | 0.09 | +11 ±5 | Sakai et al (2019) |
| G192.60-0.04 | 192.6 | -0.0 | 1.52 [1.43-1.61] | ±0.09 | 0.04 | +6 ±5 | Choi et al (2014) |
| G211.59+1.05 | 211.6 | +1.1 | 4.39 [4.26-4.52] | ±0.13 | 0.03 | +45 ±5 | Reid et al (2014) |
| G229.57+0.15 | 229.6 | +0.2 | 4.59 [4.35-4.85] | ±0.25 | 0.06 | +47 ±10 | Choi et al (2014) |
| G236.81+1.98 | 236.8 | +2.0 | 3.07 [2.84-3.33] | ±0.25 | 0.08 | +43 ±7 | Choi et al (2014) |
| G240.31+0.07 | 240.3 | +0.1 | 5.32 [4.90-5.81] | ±0.45 | 0.09 | +67 ±5 | Choi et al (2014) |

Note 1: Excluding nearby sources, identified as located in the 'local arm' or spur - see Reid et al (2014), and excluding sources with a large fractional parallax error > 0.2. When needed, the published parallax (p, in mas) was converted to a distance (D, in kpc) through the equation D = 1/p.

=================================================================

## Table 3. Trigonometric masers near the Sagittarius spiral arm (green curve).

| Name | Gal. Long. (o) | Gal. Lat. (o) | Distance D [range] (kpc) | Half range (kpc) | Fract. parall. error | Syst. $V_{lsr}$ (km/s) | Reference |
|---|---|---|---|---|---|---|---|
| G351.44+00.65 | 351.4 | +0.6 | 1.34 [1.22-1.49] | ±0.14 | 0.10 | -8 ±3 | Reid et al (2014) |
| G011.49-01.48 | 011.5 | -1.5 | 1.25 [1.20-1.30] | ±0.05 | 0.04 | +11 ±3 | Reid et al (2014) |
| G014.33-00.64 | 014.3 | -0.6 | 1.12 [1.01-1.26] | ±0.13 | 0.11 | +22 ±5 | Reid et al (2014) |
| G014.63-00.57 | 014.6 | -0.6 | 1.83 [1.76-1.91] | ±0.07 | 0.04 | +19 ±5 | Reid et al (2014) |
| G015.03-00.67 | 015.0 | -0.7 | 1.98 [1.86-2.12] | ±0.13 | 0.07 | +22 ±3 | Reid et al (2014) |
| G032.74-00.07 | 032.7 | -0.1 | 7.94 [7.04-9.09] | ±1.03 | 0.13 | +37 ±10 | Wu et al (2019) |
| G034.39+00.22 | 034.4 | +0.2 | 1.56 [1.44-1.68] | ±0.12 | 0.08 | +57 ±5 | Reid et al (2014) |
| G035.02+00.34 | 035.0 | +0.3 | 2.33 [2.13-2.56] | ±0.22 | 0.09 | +52 ±5 | Reid et al (2014) |
| G035.19-00.74 | 035.2 | -0.7 | 2.19 [2.00-2.43] | ±0.22 | 0.01 | +30 ±7 | Reid et al (2014) |
| G035.20-01.73 | 035.2 | -1.7 | 3.27 [2.85-3.83] | ±0.49 | 0.15 | +42 ±3 | Reid et al (2014) |
| G035.20-01.73 | 035.2 | -1.7 | 2.43 [2.35-2.51] | ±0.08 | 0.03 | +43 ±5 | Wu et al (2019) |
| G035.79-00.17 | 035.8 | -0.2 | 8.85 [7.94-10.00] | ±1.03 | 0.12 | +61 ±5 | Wu et al (2019) |
| G037.43+01.51 | 037.4 | +1.5 | 1.88 [1.81-1.96] | ±0.08 | 0.04 | +41 ±3 | Reid et al (2014) |
| G043.03-00.45 | 043.0 | -0.4 | 7.69 [6.71-9.00] | ±1.15 | 0.15 | +56 ±3 | Wu et al (2019) |
| G043.79-00.12 | 043.8 | -0.1 | 6.03 [5.68-6.41] | ±0.36 | 0.06 | +44 ±10 | Reid et al (2014) |
| G043.89-00.78 | 043.9 | -0.8 | 7.46 [6.80-8.26] | ±0.72 | 0.10 | +50 ±3 | Wu et al (2019) |
| G045.07+00.13 | 045.1 | +0.1 | 8.00 [7.69-8.33] | ±0.32 | 0.04 | +59 ±5 | Reid et al (2014) |
| G045.07+00.13 | 045.1 | +0.1 | 7.75 [7.35-8.20] | ±0.42 | 0.05 | +59 ±5 | Wu et al (2014) |

| Name | Gal. Long. (o) | Gal. Lat. (o) | Distance D [range] (kpc) | Half range (kpc) | Fract. parall. error | Syst. $V_{lsr}$ (km/s) | Reference |
|---|---|---|---|---|---|---|---|
| G048.99-00.30 | 049.0 | -0.3 | 5.62 [5.13-6.21] | ±0.54 | 0.10 | +67 ±1 | Nagayama et al (2015) |
| G049.19-00.34 | 049.2 | -0.3 | 4.74 [4.40-5.13] | ±0.36 | 0.08 | +70 ±1 | Nagayama et al (2015) |
| G049.19-00.33 | 049.2 | -0.3 | 5.29 [5.10-5.49] | ±0.19 | 0.04 | +67 ±5 | Reid et al (2014) |
| G049.34+00.41 | 049.3 | +0.4 | 4.15 [3.68-4.76] | ±0.54 | 0.13 | +68 ±5 | Wu et al (2019) |
| G049.48-00.38 | 049.5 | -0.4 | 5.40 [5.13-5.71] | ±0.29 | 0.05 | +58 ±4 | Reid et al (2014) |
| G049.59-00.24 | 049.6 | -0.2 | 4.59 [4.41-4.78] | ±0.19 | 0.04 | +63 ±5 | Wu et al (2019) |
| G052.10+01.04 | 052.1 | +1.0 | 6.06 [5.62-6.58] | ±0.48 | 0.08 | +42 ±5 | Wu et al (2019) |

Note : Taking all trigonometric maser sources, identified as located in the 'Sagittarius' (Carina) arm in Reid et al (2014) and Wu et al (2019), and some others, excluding sources with a large fractional parallax error > 0.2. All distances are trigonometric; the published parallax (p, in mas) was converted to a distance (D, in kpc) through the equation D = 1/p.

= = = = = = = = = = = = = = = = = = = = = = = = = = = = = = = = = = = = = = = = = = = = = = = = = = = = = = = = = = = = =

## Table 4. Trigonometric masers near the Scutum spiral arm (blue curve).

| Name | Gal. Long. (o) | Gal. Lat. (o) | Distance D [range] (kpc) | Half range (kpc) | Fract. parall. error | Syst. $V_{lsr}$ (km/s) | Reference |
|---|---|---|---|---|---|---|---|
| G005.88-00.39 | 005.9 | -0.4 | 2.99 [2.82-3.18] | ±0.18 | 0.06 | +9 ±3 | Reid et al (2014) |
| G011.91-00.61 | 011.9 | -0.6 | 3.37 [3.05-3.76] | ±0.35 | 0.10 | +37 ±5 | Reid et al (2014) |
| G012.80-00.20 | 012.8 | -0.2 | 2.92 [2.63-3.27] | ±0.32 | 0.11 | +34 ±5 | Reid et al (2014) |
| G013.87+00.28 | 013.9 | +0.3 | 3.94 [3.60-4.35] | ±0.38 | 0.09 | +48 ±10 | Reid et al (2014) |
| G016.58-00.05 | 016.6 | -0.1 | 3.58 [3.31-3.91] | ±0.30 | 0.08 | +60 ±5 | Reid et al (2014) |
| G023.00-00.41 | 023.0 | -0.4 | 4.59 [4.26-4.98] | ±0.36 | 0.08 | +80 ±3 | Reid et al (2014) |
| G023.44-00.18 | 023.4 | -0.2 | 5.88 [4.95-7.25] | ±1.15 | 0.19 | +97 ±3 | Reid et al (2014) |
| G023.70-00.19 | 023.7 | -0.2 | 6.21 [5.40-7.30] | ±0.95 | 0.15 | +73 ±5 | Reid et al (2014) |
| G028.86+00.06 | 028.9 | +0.1 | 7.41 [6.54-8.55] | ±1.10 | 0.13 | +100 ±10 | Reid et al (2014) |
| G029.86-00.04 | 029.9 | -0.0 | 6.21 [5.52-7.09] | ±0.78 | 0.12 | +100 ±3 | Reid et al (2014) |
| G029.95-00.01 | 030.0 | -0.0 | 5.26 [4.78-5.85] | ±0.53 | 0.10 | +98 ±3 | Reid et al (2014) |
| G031.28+00.06 | 031.3 | +0.1 | 4.27 [3.66-5.13] | ±0.73 | 0.17 | +109 ±3 | Reid et al (2014) |
| G031.58+00.07 | 031.6 | +0.1 | 4.90 [4.27-5.75] | ±0.74 | 0.15 | +96 ±5 | Reid et al (2014) |
| G032.04+00.05 | 032.0 | +0.1 | 5.18 [4.97-5.40] | ±0.22 | 0.04 | +97 ±5 | Reid et al (2014) |
| G348.70-01.04 | 348.7 | -1.0 | 3.38 [3.10-3.70] | ±0.30 | 0.09 | -7 ±6 | Reid et al (2014) |

Note : Taking all trigonometric maser sources, identified as located in the 'Scutum' arm in Reid et al (2014), excluding 3 sources with a nearby solar distance < 2.6 kpc (G12.68; G12.88; G12.90), and excluding sources with a large fractional parallax error > 0.2. All distances are trigonometric; the published parallax (p, in mas) was converted to a distance (D, in kpc) through the equation D = 1/p.

=============================================================

## Table 5. Trigonometric masers near the Norma spiral arm (red curve).

| Name | Gal. Long. (o) | Gal. Lat. (o) | Distance D [range] (kpc) | Half range (kpc) | Fract. parall. error | Syst. $V_{lsr}$ (km/s) | Reference |
|---|---|---|---|---|---|---|---|
| G009.62+00.19 | 009.6 | +0.2 | 5.15 [4.61-5.85] | ±0.62 | 0.12 | +2 ±3 | Reid et al (2014) |
| G010.62-00.38 | 010.6 | -0.4 | 4.95 [4.52-5.46] | ±0.47 | 0.09 | -3 ±5 | Reid et al (2014) |
| G012.02-00.03 | 012.0 | -0.0 | 9.43 [8.78-10.20] | ±0.71 | 0.08 | +108 ±5 | Reid et al (2014) |

Note : Taking all trigonometric maser sources, identified as located in the 'Norma' (4 – k) arm and 'Inner Perseus' (3 –k) arm in Reid et al (2014), and some others, excluding sources with a large fractional parallax error > 0.2. All distances are trigonometric; the published parallax (p, in mas) was converted to a distance (D, in kpc) through the equation D = 1/p.

=============================================================

## Table 6. Trigonometric masers near the Outer Scutum spiral arm (blue curve).

| Name | Gal. Long. (o) | Gal. Lat. (o) | Distance D [range] (kpc) | Half range (kpc) | Fract. parall. error | Syst. $V_{lsr}$ (km/s) | Reference |
|---|---|---|---|---|---|---|---|
| G007.47+00.05 | 007.5 | +0.1 | 20.4 [18.18-23.26] | ±2.54 | 0.12 | -16 ±4 | Sanna et al (2017) |

Note : Taking a trigonometric maser source, identified as located in the 'Cygnus +I' (outer Scutum) arm in Sanna et al (2017), and some others, excluding sources with a large fractional parallax error > 0.2. All distances are trigonometric; the published parallax (p, in mas) was converted to a distance (D, in kpc) through the equation D = 1/p.

= = = = = = = = = = = = = = = = = = = = = = = = = = = = = = = = = = = = = = = = = = = = = = = = = = = = = = = =

**Table 7 – Offsets of maser locations, from the CO-peaks (positive, towards the Galactic Center).**

| Arm name | Galactic Quadrants | Median[1] offset (kpc) |
|---|---|---|
| Cygnus (Outer Norma) | II and III | 1.25 ± 0.06 |
| Perseus arm | II and III | 0.31 ± 0.06 |
| Sagittarius | I and IV | 0.38 ± 0.06 |
| Scutum | I and IV | 0.19 ± 0.06 |
| Norma | I | 0.12 ± 0.06 |
| Median | - | 0.34 ± 0.06 |

Note 1: Median of data from Figure 1 here. Positive values are located toward the Galactic Center. Each offset (CO to maser) is measured in Figure 1b, taking the shortest distance of a maser to the nearest model arm, with a separation error of 0.06 kpc.

= = = = = = = = = = = = = = = = = = = = = = = = = = = = = = = = = = = = = = = = = = = = = = = = = = = = = = = =

## Table 8 – Offsets of maser velocities, from the 4-arm CO model (positive, away from the Sun).

| Arm name | Galactic Quadrant | Galactic longitude (o) | Maser $V_{lsr}$ (km/s) | Offset[3] from CO model (km/s) |
|---|---|---|---|---|
| Cygnus (Outer Norma) | II | 130 – 200 | individual[1] | -8 ± 1.3 |
| Perseus arm | II | 100 - 140 | -49 ± 3 [2] | -10 ± 0.6 |
| Sagittarius | I | 035 - 046 | +49 ± 3 [2] | +29 ± 0.6 |
| Scutum | I | 022 - 030 | +90 ± 4 [2] | +10 ± 0.8 |
| Median (absolute value) | - | - | | 10 ± 0.8 |

Note 1: Data from Figure 2b here. Each offset is measured by the shortest vertical distance of a maser to the nearest model arm, with a separation error of 1.1 km/s.

Note 2: Observed data from Tables 2 (Perseus), 3 (Sagittarius), and 4 (Scutum) here. Each offset is measured by the shortest vertical distance of a maser to the nearest model arm, in Figure 2b (Perseus) of 2a (Sagittarius, Scutum, Norma), with a separation error of ±1.1 km/s. The combined individual error, including the maser error at ±3 km/s, is quadratically added, giving ±3.2 km/s.
The statistical error of the mean is that divided by the square root of the number of masers

Note 3: Positive values for sources moving away from the Sun.
= = = = = = = = = = = = = = = = = = = = = = = = = = = = = = = = = = = = = = = = = = = = = = = = = = = = = = = =

**Figure Captions**

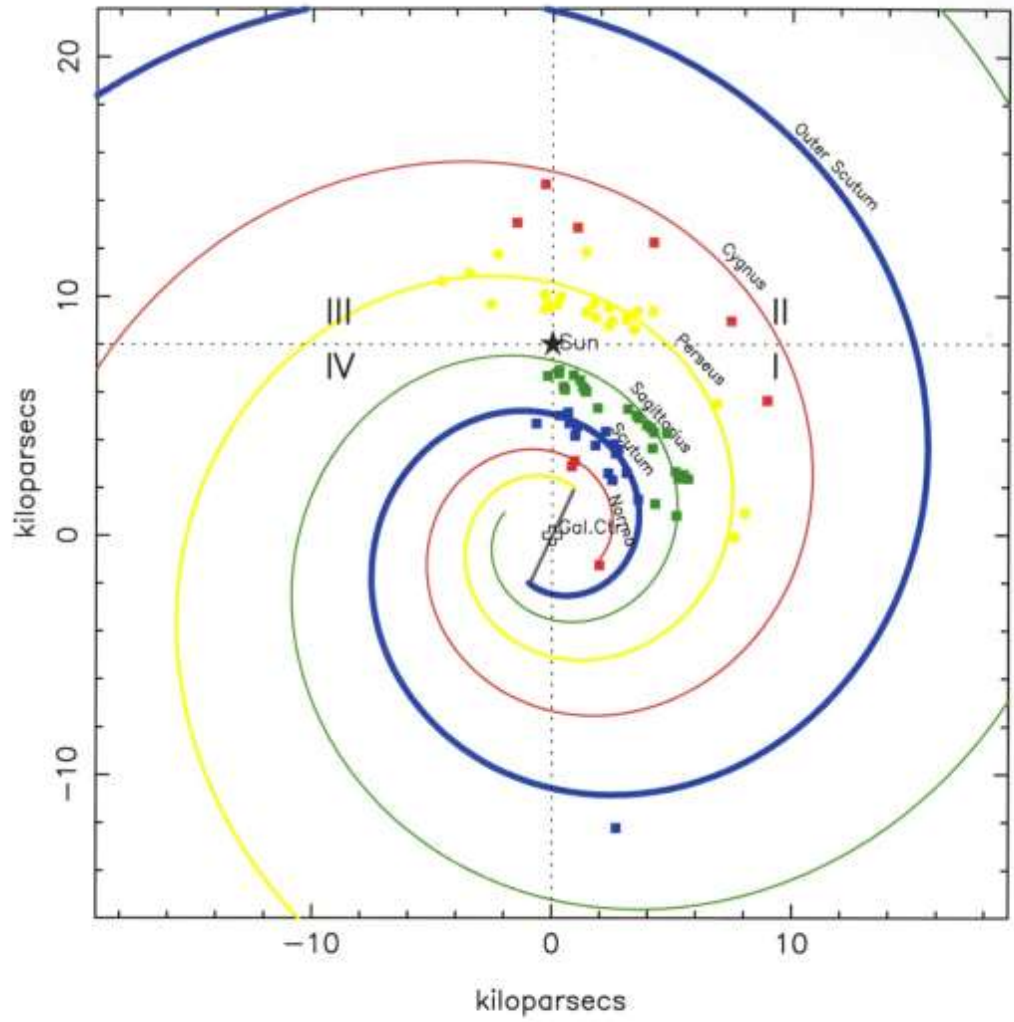

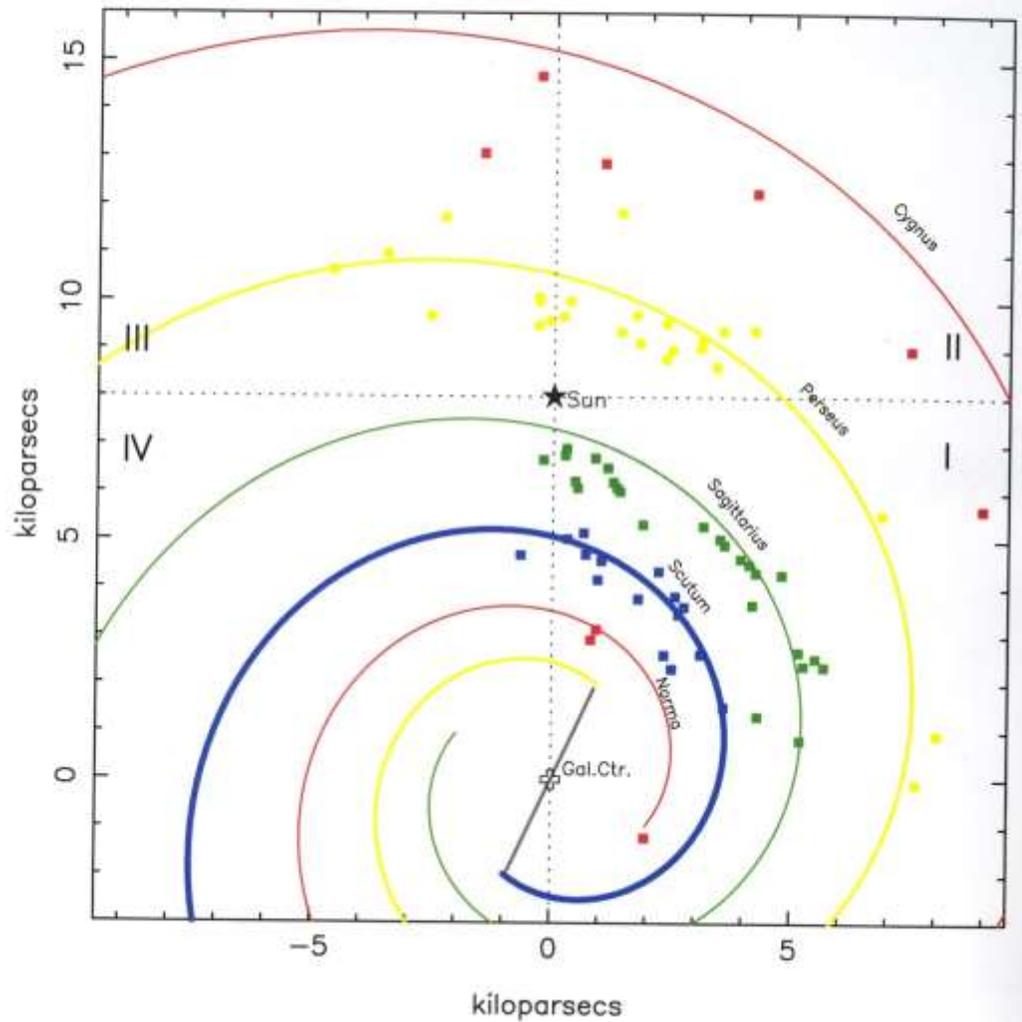

**Figure 1.** Locations of the masers in the galactic disk, as seen from above the disk. The Sun is at 8.0 kpc, north of the Galactic Center (0,0). Galactic Quadrants I to IV are shown, along with the 4 spiral arms. The small error bars for each maser are indicated in Tables 1 to 6.
  (a) A wide view is shown, reaching 22 kpc at the top.
  (b) A narrow view is shown, showing the masers being preferentially on the inner side of their spiral arms (closest to the Galactic Center).

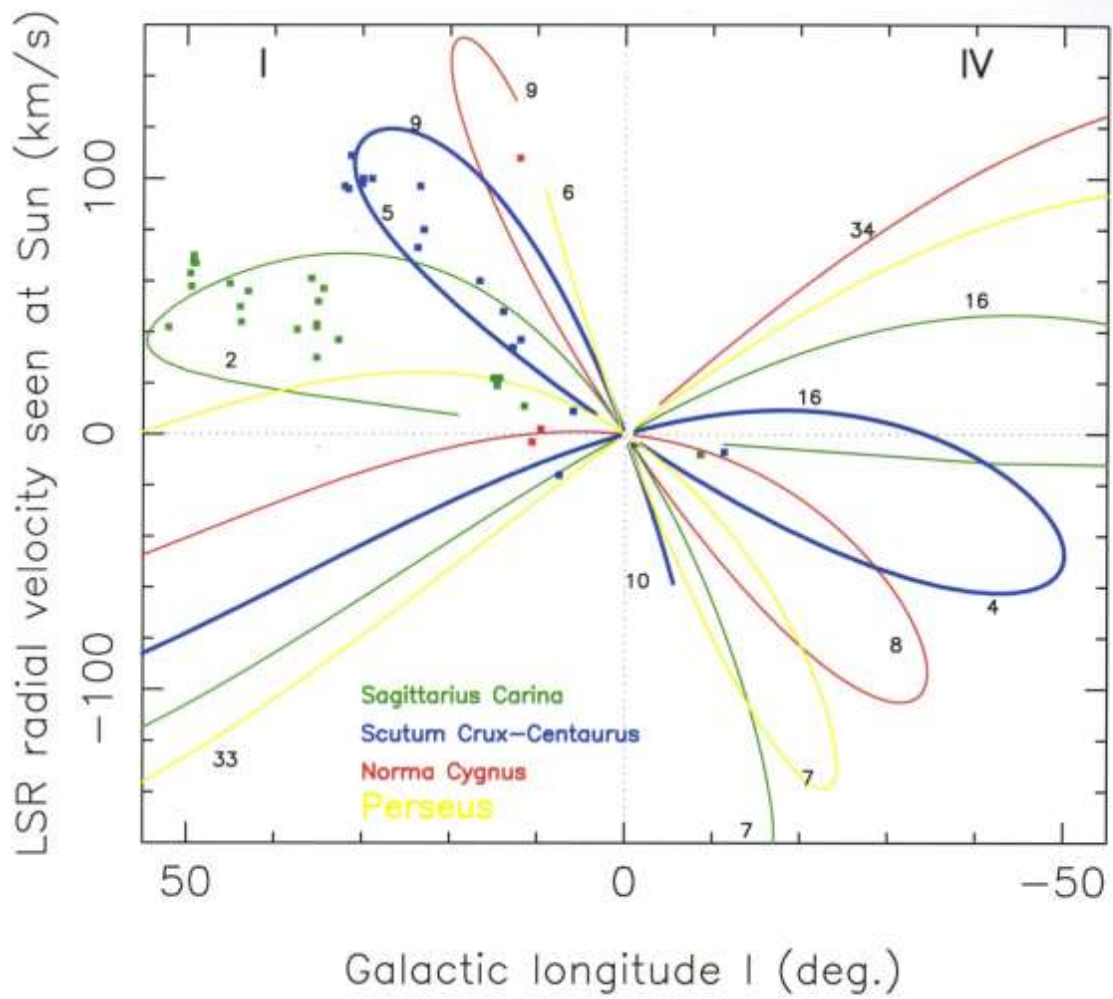

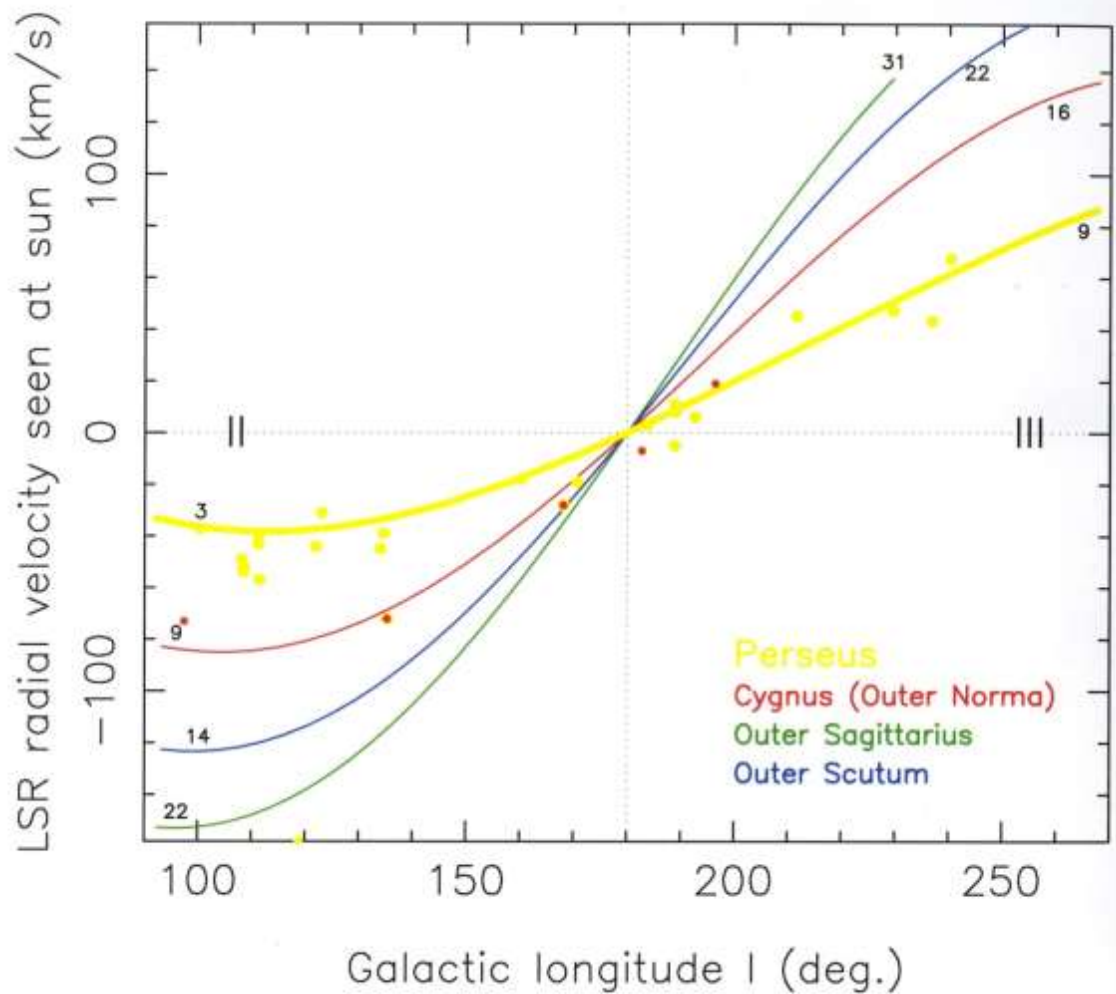

**Figure 2.** Locations of masers in velocity space (vertical axis) and galactic longitudes (horizontal axis). The spiral arms are shown as curves. Numbers on the arms indicate the rough distance of that arm segment to the Sun. The small error bars for each maser are indicated in Tables 1 to 6 (column 7).
(a) Here the masers (squares) are shown towards the general area of the Galactic Center (longitudes -90 to +90 degrees). The gas beyond the Galactic Center are shown by the curves at bottom left (Galactic Quadrant I, at negative velocities) and the curves at top right (Galactic quadrant IV, at positive velocities).
(b) Here the masers (squares) are shown towards the general area of the anti-galactic center (longitudes 90 to 270 degrees).

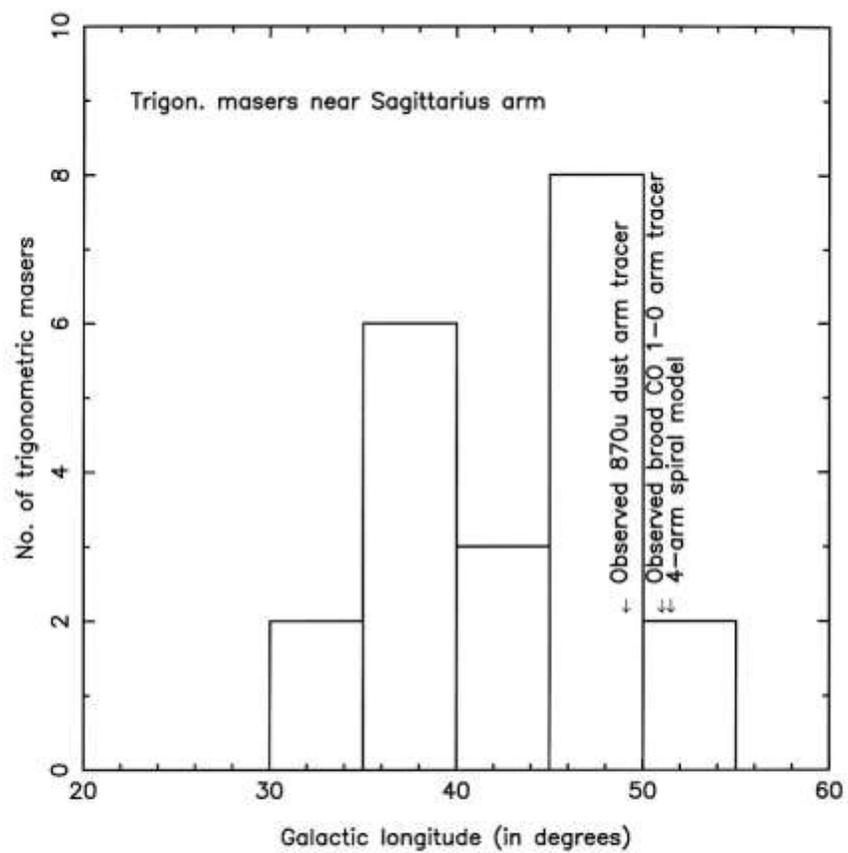

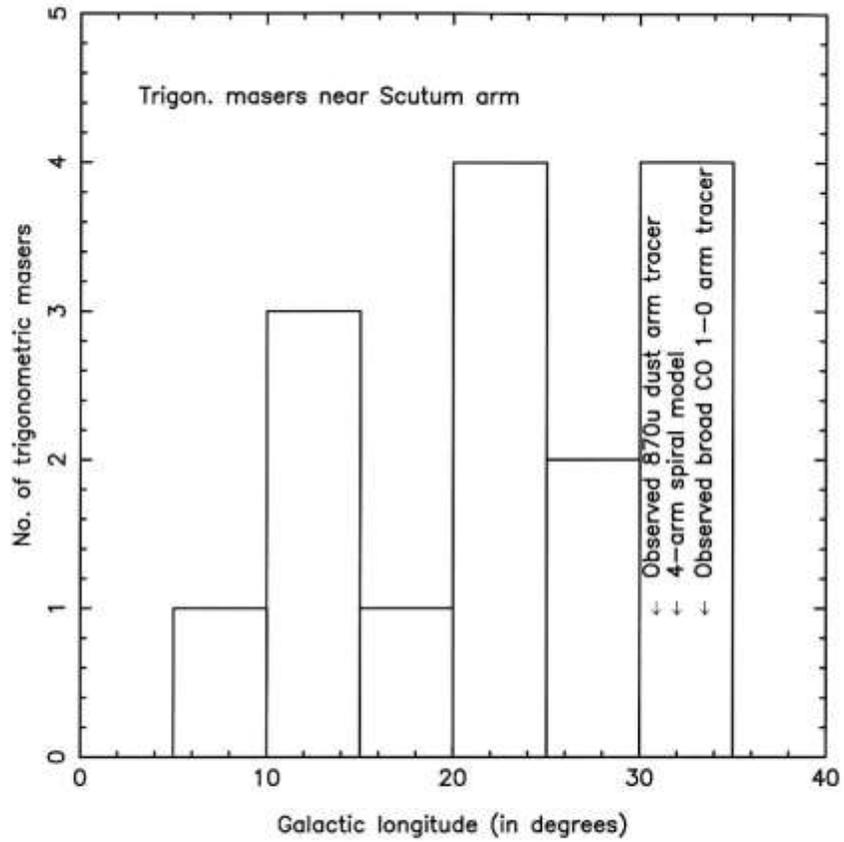

**Figure 3.** Histograms of the locations of the trigonometric masers, in galactic longitudes. The location of the median longitude for the observed broad CO J-1-0 arm tracer is shown, as well as the longitudes of the observed 870μ dust arm tracer, and the fitted 4-arm model.
    a) Near the Sagittarius arm.
    b) Near the Scutum arm.

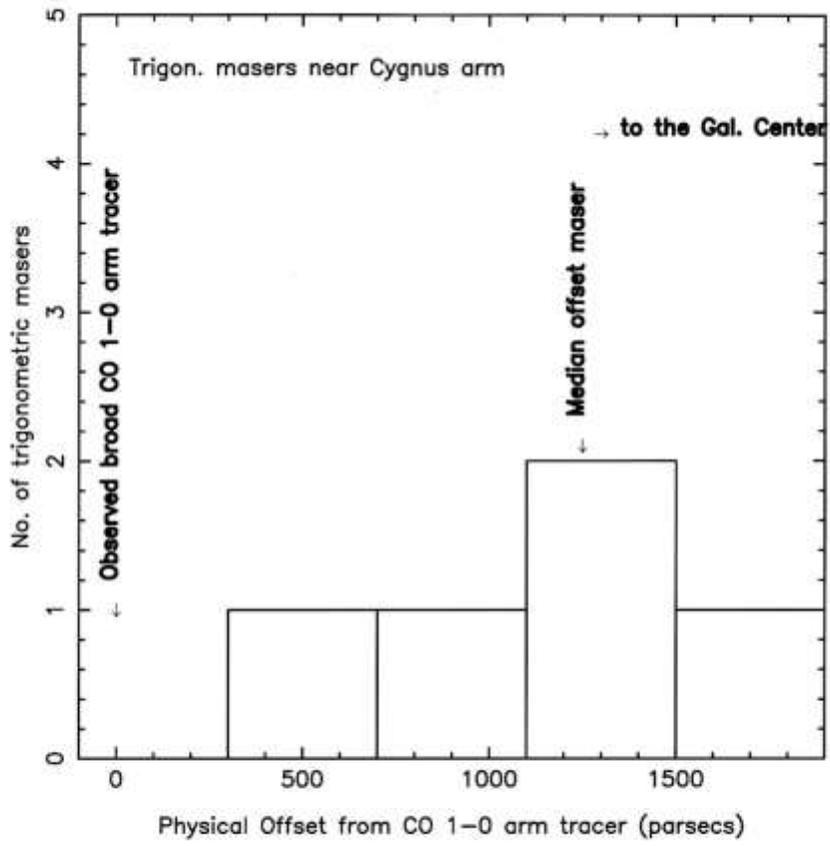

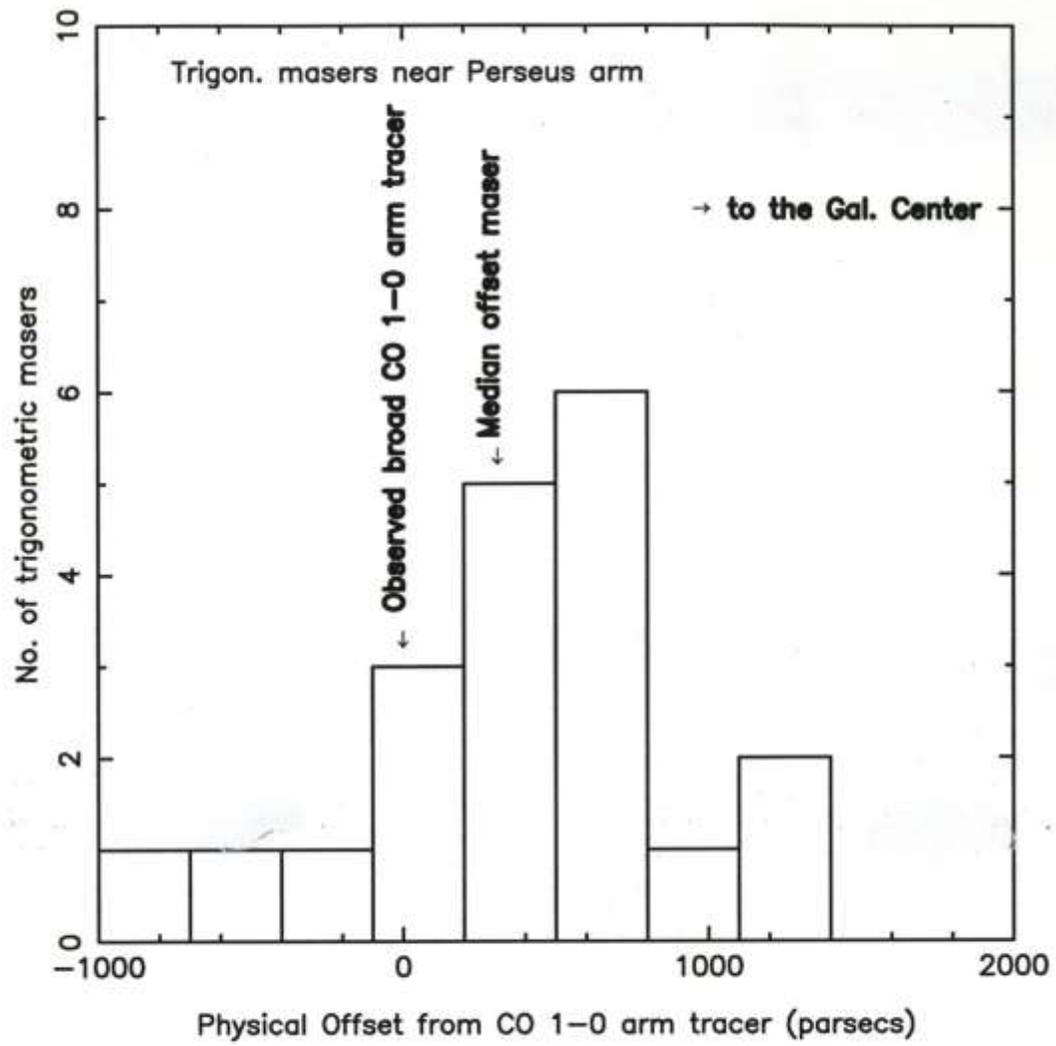

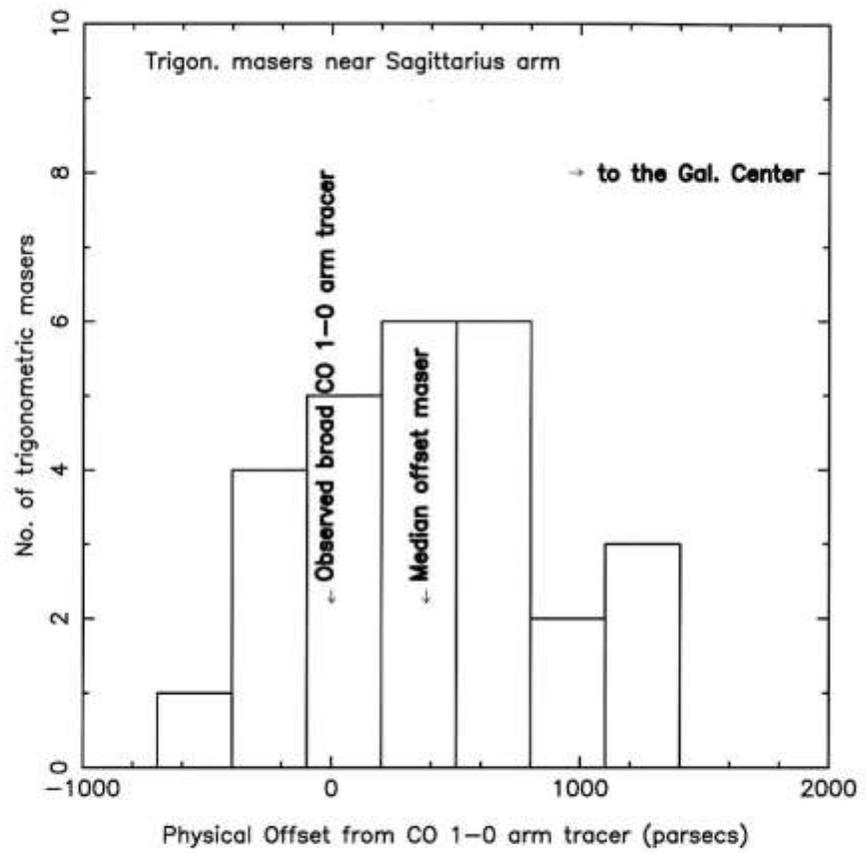

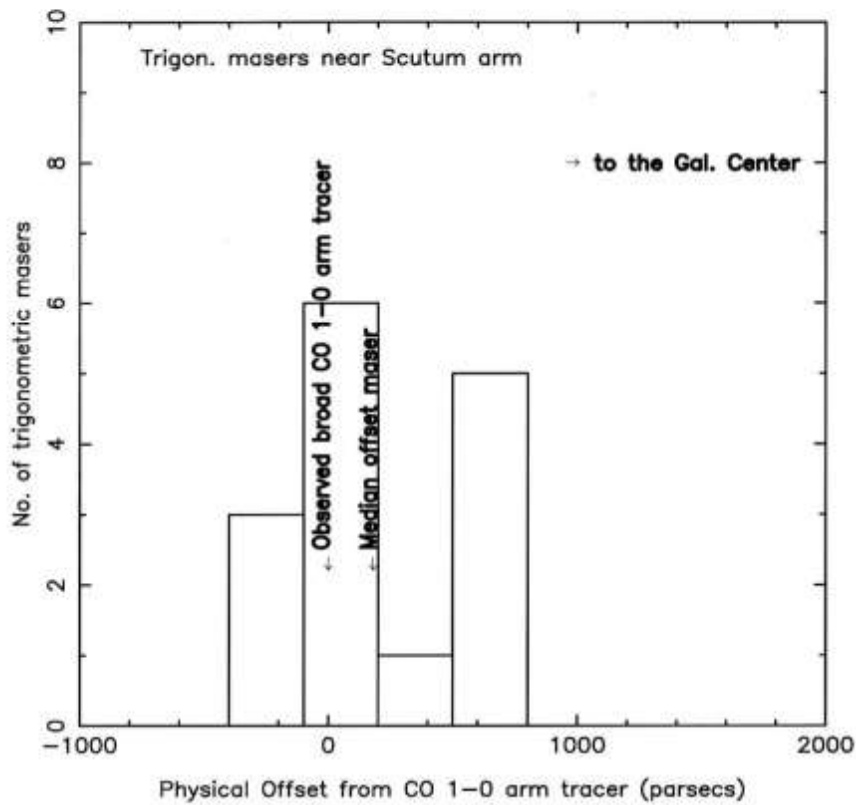

Figure 4. Histograms of the maser locations, as a physical offset from the nearest CO 1-0 gas arm tangent (tabulated in Table 5 in Vallée 2015, or its model projection).
 The mid-arm is taken as the location of the old stars and the cold low-density CO 1-0 gas, seen where a scan in galactic longitude shows a peak in CO 1-0 emissivity due to looking at an arm tangent (along the long line of sight). A subsequent fit of a 4-arm spiral is done to match the specific galactic longitude of the maximum CO 1-0 emissivity for each arm. As found elsewhere, that galactic longitude of maximum emissivity for CO 1-0 is always farther away from the galactic longitude of that arm's dust lane (away from the Galactic Center direction).
 a) Near the Cygnus arm.
 b) Near the Perseus arm.
 c) Near the Sagittarius arm.
 d) Near the Scutum arm.